# Nanoreplicated positive and inverted sub-micron polymer pyramids array for surface enhanced Raman spectroscopy (SERS)

Zhida Xu[1*], Hsin-Yu Wu[1], Syed Usman Ali[1], Jing Jiang[1], Brian T Cunningham[1,2], Gang Logan Liu[1,2]

[1]Department of Electrical and Computer Engineering, University of Illinois at Urbana-Champaign, Urbana, 61801, USA
[2]Department of Bioengineering, University of Illinois at Urbana-Champaign, Urbana, 61801, USA
* zhidaxu1@illinois.edu

## Abstract

We demonstrate gold coated polymer surface enhanced Raman scattering (SERS) substrates with a pair of complementary structures--positive and inverted pyramids array structures fabricated by multiple-step molding and replication process. The uniform SERS enhancement factors over the entire device surfaces were measured as $7.2 \times 10^4$ for positive pyramids substrate while $1.6 \times 10^6$ for inverted pyramids substrate with Rhodamine 6G as the target analyte. Based on the optical reflection measurement and FDTD simulation result, the enhancement factor difference is attributed to plasmon resonance matching and to SERS "hot spots" distribution. With this simple, fast and versatile complementary molding process, we can produce polymer SERS substrates with extremely low cost, high throughput and high repeatability.

## 1. Introduction

Surface-enhanced Raman spectroscopy (SERS) is a highly sensitive and versatile technique in analytical chemistry. The applications of SERS span a range of fields including bacteria detection, water contamination detection, protein-DNA interaction analysis, forensic investigation and archeological identification [1-4]. The most important role in SERS is played by the roughed metal surface, which can enhance the Raman scattering of molecules adsorbed or in proximity to the surface. In terms of the material, the coinage metals including gold, silver and copper are most commonly used as enhancing metal because their plasmonic resonance in visible to near infrared range can further boost the scattering enhancement when the excitation light is in the corresponding wavelength range [5,6]. In addition to the material, the surface structure is also critical. According to the electromagnetic field enhancement theory, the enhancement factor is proportional to the fourth power of the amplitude of local electric field [5, 6]. To get maximum local field enhancement, sharp tips and closely packed particles are preferred to create "hotspots" for SERS.[7] The plasmonic properties of the metal surface are also affected by the morphology of surface.[8-10] Extensive efforts have been made on design, modeling, fabrication and characterization of Raman enhancing metal surface, which we call SERS substrate, for decades since the electrochemically roughened silver electrodes used as the first SERS substrate.[11] With the advancements in nanofabrication, different structures of SERS substrates are produced with a variety of techniques. We can roughly divide the structures into two categories-the random structure and well patterned structure. The random types include colloid metal particles cluster, nanopillar made with anodized aluminum oxide (AAO) template and glancing angle deposition, dark silver produced by plasma etching and so on[12-15]. The well patterned types include synthesized metal particle pairs and a variety of nanostructure array made with photolithography, ebeam lithography, assembling nanosphere lithography, deep reactive ion etching, soft lithography and so on.[16-20].Compared with random structures in terms of fabrication, well patterned structures are more predictable, tunable and reliable but the cost is higher and the throughput is much lower. The plastic replica molding provides a solution for the weakness of well patterned structures as it can promote replication of nanoscale structures over large areas in a rapid and cost-effective manner and potentially facilitate their integration in lab-on-a-chip devices [19].

The idea of making SERS substrates by plastic molding has been proposed for years and some efforts have been made on producing them [21,16,19,22]. However, the limitations hindering its broad application were soon realized. Firstly, the plastic itself will give Raman peaks which will interfere with the target analyte. Secondly, as the plastic may be easily melted by the focused laser beam, which is to excite the Raman signal, the nanostructure will be damaged. Thirdly, the templates for molding, which are usually textured silicon wafer, can not be used for infinite times. The silicon template is usually difficult and expensive to fabricate and will wear away after certain times of molding. But even worse, most templates are scrapped when plastics permanently sticks to the silicon template and it is hard to remove the residue plastics without breaking the nanostructure. To overcome these problems, we propose a complementary templated molding process (CTM). Instead of using silicon wafer as the molding template, we use the plastic replica itself as the molding template to make its complementary replica. Then we use this complementary replica as molding template to make the replica with original structure again. In this way, the silicon template was used only once at the beginning. Considering that the inverted pyramids array structure has been demonstrated to have uniform enhancement factor (EF) about $10^6$ [23, 24] and is very suitable for replication, we choose it as the original molding template to make positive pyramids replica. Then we can make complementary inverted pyramids replica by molding the positive pyramids replica. Since the inverted pyramids replica has the same structure as the gold coated silicon substrate, it should give same optical properties and SERS performance as well. To avoid the melting problem, we choose ultraviolet (UV) curable polymer instead of thermal cure polymer as substrate plastic. To avoid the interference from the polymer Raman signal, we make sure the metal coating is thick enough to prevent light from reaching the plastic, which also works as heat sink due to the excellent thermal conductivity of gold.

To investigate the plasmonic properties of the substrates, we take reflection spectra measurement and finite element time domain (FDTD) simulation for the positive and inverted replica polymer SERS substrate and silicon SERS substrate. Both measurement and simulation show the reflection spectra for inverted pyramids polymer replica and silicon substrate have a dip around 785nm, which is the excitation wavelength we use for SERS, while not seen for the positive pyramids replica. This is predictable since the gold-coated inverted silicon pyramid substrate is intentionally designed to be resonant at 785nm [23, 24]. Accordingly, in SERS measurement, both inverted pyramids replica and the original silicon substrate, in resonance with excitation, show the EF of $1.6 \times 10^6$ while the positive pyramids, off resonance, shows much smaller EF of $3.2 \times 10^4$. However, this is not to say the positive pyramids replica is worse than inverted pyramids replica in SERS but to say under this geometry the inverted pyramids replica has the resonant modes near 785nm. Actually positive pyramids replica may have higher field enhancement at other excitation wavelengths due to the sharp tip of the pyramid if tuned to be in resonance [21, 25].

## 2. Experimental details

2.1 Materials

To remove the gold coating, TFA gold etchant (Transene Inc.) containing 8 wt% Iodine, 21 wt% Potassium Iodide, 71 wt% water, with etching rate of 28Å per second was used. To make the silicon surface hydrophobic, Real-Silane (PlusOne, GE healthcare inc.) was used. Rhodamine 6G (Sigma Aldrich) was used as Raman target analyte. The UV curable polymer is acrylate modified silicone polymer with curing spectral range of 250–364 nm (Zipcone, Gelest Inc.)

2.2 Instrumentation

To cure the UV curable polymer, a high intensity pulsed UV curing system (Xenon Inc.) with spectral cutoff at 240 nm, peak power density of 405 W cm−2, pulse repetition rate of 120 Hz and pulse width of 25 μs was used. A six pocket E-Beam Evaporation System (Temescal) was used for metal evaporation. For reflection spectra measurement, a micro-spectroscopy workstation built upon a Zeiss Axio Observer D1 inverted microscope was used. Scanning Electron Micrograph (SEM) images were taken with Hitachi S4700 SEM system.

A home made back scattering setup was used for Raman spectroscopy measurement, shown in Fig. 1S. A semiconductor laser with the wavelength of 785nm and power of 30mW was used for excitation. The excitation light is delivered and back scattered light is collected both by a 10X 0.28NA objective lens (Mitutoyo Inc.). The Raman scattering light is directed to a spectrometer (Acton Inc.) through a series of mirrors and lens and the laser light and Rayleigh scattering light was filtered out by a dichroic filter and a long-pass filter.

2.3 Positive and inverted SERS substrates fabrication

Figure 1 is the cross section sketch of fabrication process of positive and inverted replica SERS substrates. All the replication processes are carried on using low forces at room temperature to produce large area of uniform sub-micron structure in several minutes. First of all, we obtained the inverted pyramids silicon substrate by removing the gold coating from silicon SERS substrate with gold etcher for ten minutes at room temperature. Figure 2(a) shows the photograph of original gold coated silicon SERS substrate while figure 2(b) shows that after gold removed. This is the primitive template. The well-known method of creating inverted pyramids array on silicon is potassium hydroxide (KOH) anisotropic etching following photolithography, the process of which is shown in Fig. 2S. [23, 26] The process of producing positive pyramids replica is shown in Fig. 1(a-c). Firstly, the silicon mold template was immersed in dimethyl dichlorosilane solution for 5 min followed by ethanol and DI water rinse. This treatment creates a hydrophobic silane layer on the silicon template surface which prevents cured polymer replica from adhering and therefore promotes clean release of the replica. Then a 250 μm thick flexible polyethylene terephthalate (PET) sheet was placed on top for peeling off later on, and a Teflon roller was used to press and distribute the liquid polymer layer evenly between the silicon mold and the PET sheet. The liquid polymer which conformed to the shape of the features on the wafer was subsequently cured to solid state after being exposed to UV light for 90 seconds. After curing, the molded structure was released from the wafer by peeling away the PET, resulting in a polymer complementary replica of the silicon substrate adhered to the PET sheet, which is the positive pyramids array replica (Fig. 3S(a)(b)). With this replica as the molding template, we produced the inverted pyramids array using the same replication processes on silicon template, shown in Fig. 1(d-f). If we directly mold the positive pyramids replica with the same polymer, they are likely to adhere together thus difficult to take apart. To circumvent this problem, before the replication we deposited a layer of silicon dioxide with thickness of 20nm onto the positive pyramids replica then treated it with dimethyl dichlorosilane solution to make it hydrophobic as we did for silicon substrate. The inverted pyramids replica turned out to be separable from the positive pyramids replica. In the same way, we managed to produce positive pyramids replica again with the inverted pyramids replica as molding template (Fig. 3S(c)(d)). So far we demonstrated two kinds of replica which can work as templates for each other and thus we call these processes complementary templated molding (CTM). The silicon template is not needed for molding process after it was used only once at the beginning. The silicon template is usually difficult and expensive to fabricate and will wear away after certain times of molding. But even worse, most templates are scrapped when a portion of plastic get sticked to the template thus hard to remove without breaking the nanostructure. With CTM, we are able overcome this problem.

For the plasmonic enhancement of Raman scattering, we need to coat the replica substrate with coinage metal. We chose to deposit gold with electron beam evaporation due to its resistance to oxidation compared with silver and copper. To promote the adhesion, we first deposited 10nm of Titanium with evaporation rate of 0.05nm/s on to the polymer substrates followed by deposition of 200nm of gold with evaporation rate of 0.5nm/s. Concerning about several issues, we intentionally evaporated gold with extensive thickness. The first thing we consider is to avoid interference from the polymer. If the gold is too thin the light can transmit to reach the polymer underneath and excite Raman scattering of polymer which may interfere with the Raman spectra of target analytes. The second issue need to be considered is the photothermal effect. When laser beam is focused on the SERS substrate the heat may be accumulated to melt the polymer. Gold also works as heat sink here due to its excellent thermal conductivity. The SERS substrates are completed after gold deposition. Figure 2(c) is the photograph of the positive pyramids replica SERS substrate and Figure 3(a) and (b) are SEM images. Figure 2(d) is the photograph of the inverted pyramids replica SERS substrate and Figure 3 (c) and (d) are SEM images.

2.4 Electromagnetic modeling

To calculate the electromagnetic field distribution in the fabricated SERS substrate numerically, finite difference time domain (FDTD) method was used to solve Maxwell's equations in three dimensions in the solution domains (air, gold and UV curable polymer). Both total electric field and magnetic field are composed of incident field and scattered field: $E=E_{in}+E_{sc}$ and $H=H_{in}+H_{sc}$. The incident field is a plane wave defined by ourselves. After solving the total electric and magnetic field, we got the scattered field by subtracting the incident field from the total field. The SERS enhancement factor is described by the following equation: [5]

$$EF_{sers} \propto \left(\frac{|E_{sc}(\omega_{ex})|}{|E_{in}(\omega_{ex})|}\right)^2 \left(\frac{|E_{sc}(\omega_{Ra})|}{|E_{in}(\omega_{Ra})|}\right)^2 \approx E^4 \tag{1}$$

where $E_{sc}(\omega_{ex})$ is the amplitude of the enhanced local scattered electric field at the laser excitation frequency, $E_{in}(\omega_{ex})$ is the amplitude of the incident electric field at the laser excitation frequency, $E_{sc}(\omega_{Ra})$ is the amplitude of the enhanced local scattered electric field at the Raman scattered frequency, and $E_{in}(\omega_{Ra})$ is the amplitude of the electric field at the Raman scattered frequency. Since the Raman frequency shift is small compared to incident wave frequency, the enhancement factor is approximately proportional to the 4$^{th}$ power of the amplitude enhancement of incident electric field, $(E_{sc}/E_{in})^4$.

In our 3D model, the simulation domain is within a cubic. The dimensions were measured with SEM (Fig. 4S). The base lateral length of both silicon positive and inverse pyramids is 3 μm, the spacing distance of adjacent pyramids in both x and y direction is 3.8 μm. The angle between the flat plane (100) and inclined surface (111) is 54.7° due to KOH anisotropic etching and the height of pyramid is 2.1 μm (sketched in Fig. 4S). A gold layer with 200nm thickness is covering on silicon pyramids. The surrounding media is air. The optical constants (real and imaginary part of refractive index) of each material at wavelength range from 400 nm to 900 nm are obtained by polynomial fitting of data in Palik handbook (Fig 5S) [27].To simulate the periodicity, period boundary condition is used on four side walls of the simulation cubic domain. From the top view as shown in Fig. 4S, there is one entire pyramid in the center, one quarter of pyramid on each of the four corners of the simulation square area and one half of pyramid on each of the four sides. Perfect matched layer (PML) boundary condition is used on top and bottom sides of the cubic to attenuate reflection. As to the outer boundary of PML, scattering boundary condition is used to further reduce the reflection. Since there is a trade off between the computation accuracy and computation time depending on the meshing size, we let the mesh size adapt to the local refractive index or wavelength automatically that the size of a Yee cell in a certain domain is one percent of the wavelength in that domain. It took several hours to run the 3D FDTD simulation on a personal computer.

## 3. Results and discussion
### 3.1 SERS measurement result
To experimentally interrogate the Raman scattering enhancing property of our pyramids SERS substrate, rhodamine 6G (R6g) molecule was used as the target analyte. We dropped 1 μL R6g solution with different concentrations on the substrates and let it dry. Then it was excited with a laser diode with the wavelength of 785nm and the power of 30mW. The scattered light was sent to a spectrometer after the excitation light was filtered out (Fig. 7S). All the spectrums were acquired with the integration time of 5s. To make sure of the repeatability and reliability of the measured results, on each sample we took multiple measurements at different locations and averaged the spectrua. Every curve in Figure 4 is an averaged spectrum. We also took the Raman spectroscopic measurement of R6g on a gold coated inverted silicon pyramid SERS substrate and a gold coated smooth silicon wafer for reference (Fig. 7S).

Figure 4 (a) and (b) show the Raman spectra of R6g with the concentrations of 100nM, 1uM, 10uM and 100uM on positive and inverted pyramids SERS substrates respectively. The inverted pyramids SERS substrate gives higher Raman peaks than the positive pyramids SERS substrate at the same R6g concentration, especially at the concentration of 1uM where Raman peaks are hard to see for positive pyramids but still significant for inverted pyramids. The gold coated inverted silicon pyramids substrate shows almost identical Raman intensity as our gold coated inverted polymer pyramids substrate (Fig. 7S).

We precisely calculated the enhancement factors of positive and inverted pyramids substrates as well as gold coated inverted silicon pyramids substrate by comparing the intensity of R6g major characteristic Raman peak at the wavenumber of 1370cm$^{-1}$ and at the concentration of 1uM. 1mM R6g solution was deposited on smooth silicon wafer as the reference. To obtain the precise intensity of spectral peaks, we removed the fluorescence baseline using an iterative multi-polynomial fitting algorithm [28]. The experimentally measured enhancement factor is given as: [29]

$$EF_{sers} = \frac{I_{sers}/N_{surf}}{I_{ref}/N_{bulk}} \qquad (2)$$

where $I_{SERS}$ is the surface-enhanced Raman intensity of the characteristic peak, $N_{surf}$ is the number of molecules within the enhanced field (hot spot) region of the metallic substrate contributing to the measured SERS signal, $I_{ref}$ is the Raman intensity of the characteristic peak from the reference region, and $N_{bulk}$ is the number of molecules within the excitation volume of for the analyte on the reference region illuminated by the laser spot. The number of molecules N is calculated as:

$$N = \pi r^2 h c N_A \qquad (3)$$

where r is the radius of the excitation laser spot, h is the thickness of the R6G spot on the reference region, c is the molar concentration of the R6G analyte on the reference region, and $N_A$ is the Avogadro's number.

The spatially averaged Raman enhancement factor is measured and calculated as $7.2 \times 10^4$ for positive pyramids substrate while $1.6 \times 10^6$ for both gold coated inverted polymer pyramids substrate and gold coated inverted silicon pyramids substrate.

3.2 Simulation and measurement of optical reflection spectra
To answer the question why the measured SERS enhancement factor is stronger on inverted pyramids than on positive pyramids, we investigated the reflective properties of the substrates by experimental measurement and computer simulation. The optical reflection measurement was taken on a micro-spectroscopy workstation with non-polarized white light for illumination. The reflected light was collected by an objective and sent to a spectrometer for analysis. The reflection spectrum was normalized with respect to the spectrum of the illumination light source.

The red solid curve in Figure 5(a) is the measured reflection spectra on positive pyramids SERS substrate while the one in Figure 5(c) is the measured reflection spectra on inverted polymer pyramids SERS substrate. The blue solid curve in Figure 5(c) is the measured reflection spectra on the gold coated silicon pyramids SERS substrate. On Figure 5(c), we can see the reflection spectra for the inverted polymer pyramids SERS substrate (red curve) and the silicon SERS substrate (blue curve) have similar profiles and both of them have a dip around 785nm. This is predictable since the inverted pyramids substrate is a replica of the silicon SERS substrate, which is designed and optimized to be at resonance of 785nm. [23, 24] In comparison, the positive pyramids substrate does not show any dip at 785nm but shows one at 810nm, red curve in Fig. 5(a), indicating it is not in resonance at 785nm. This is one reason we propose why the SERS enhancement factor of inverted pyramids is higher than that of positive pyramids when excited with 785nm laser.

To simulate the reflection spectra in FDTD model, we sent in a normal incident plane wave from the top as a Gaussian pulse in time domain. The polarization is along the square edge. (Fig.5 and Fig. 6S)We obtained reflection spectra by normalizing the Fourier transform of reflected power through a plane above the pyramids with the Fourier transform of incident Gaussian pulse signal. The green dotted curves in Figure 5(a) and (c) represents the simulated spectra on positive and negative pyramids SERS substrates. The simulated spectra for positive pyramids shows a major dip around 810nm while the result for inverted pyramids shows the major dip around 785nm, which matches the measurement results, showing that the inverted pyramids substrate is more resonant at 785nm. Both measurement and simulation indicate the different SERS enhancement factors on positive and inverted pyramids substrates are attributed to plasmon resonance at different wavelength.

In additional to the plasmon resonance, we propose another reason for the enhancement factor difference by looking at the electric field distribution in simulation. Figure 5(b) and (d) show vertical cross-section of normalized local electric field distribution on positive pyramids and negative pyramids SERS substrate simulated in 3D FDTD model at the excitation of 785nm. From Figure 5(b) we can see for positive pyramids, the electric field is enhanced more, and plasmonic energy is concentrated more on the tips of the pyramids and the region in between two adjacent pyramids. While shown in Figure 5(d), for inverted pyramids, the electric field is more enhanced on the side wall of the inverted pyramidal well near the pit. That can be another reason why the inverted pyramids substrate works better for SERS. Intuitively, after the R6g solution is deposited on the inverted pyramids substrate and let dry, the molecules are more likely to stay on the side wall of the pyramidal well, especially near the pit, where the electric field is enhanced most according to the simulation. While for positive pyramids substrate, very few molecules will stay on the tip of pyramids, not to mention to be suspended in the region between pyramids. Actually the molecules are more likely to stay on the side wall of pyramids and the flat spacing surface between the pyramids, where the field enhancement is very weak according to the simulation. The stronger Raman enhancement of inverted pyramids SERS substrate is possibly due to that the "SERS hot spots" is the place where the molecules most probably reside.

### 3.3 Discussion
So far we have demonstrated the fabrication process, SERS measurement, optical reflection characterization and FDTD simulation of both positive and inverted polymer pyramids replica SERS substrates. However, there are several questions to be answered.

One question is whether inverted pyramids substrate works better than positive pyramids in SERS? The answer is not necessarily. Even though our experiments show the Raman enhancement factor of inverted pyramids substrate is much higher than that of positive pyramids substrate, it does not mean that is always the case. As we explained, one reason is the plasmon resonance matching the excitation for inverted pyramids. Since the plasmon resonance is determined by several factors including the refractive index of media, shape and size of the pyramid, the layout and periodicity of the pyramid array, we can tune the plasmon resonance by changing these factors. For instance, we can carry out SERS or reflection measurement when the substrates are immersed in different media such as oil or water for comparison. Actually, the positive pyramid has a significant advantage, its sharp tip, which is able to dramatically enhance the local field due to lighting rod effect. It was demonstrated that the sharper the tips are, the stronger the field enhancement as well as SERS are.[21] For those reasons, we believe that if plasmon resonance is tuned to be matching with excitation and in the situations when the analytes are more likely to aggregate on the top, positive pyramids may perform excellent in SERS.

Since we have demonstrated that gold coated inverted polymer pyramids substrate and the silicon substrate have almost identical enhancement factor, another question is what is good of SERS substrates made by CTM? In addition to the low cost, high throughput, simple and convenient fabrication process, the plastic substrate has some other strength. Firstly, it is easy to make two complementary replicas with respect to each other. Secondly, replica molding is so versatile that a variety of surfaces can be replicated. Above all, plastic has an overwhelming advantage, the flexibility. Its malleability and resilience allows the plastic substrate to be stretched and bended in many occasions, making it compatible and integratable with MEMS. Furthermore, the optical property may be changed by the deformation, indicating the plasmon resonance is tunable with the flexibility.

The third question is does the plastic replica SERS substrate have some limitations? It does. As mentioned before, the plastic Raman interference issue and heat melting issue are already solved by our CTM process. We observed another problem for plastic substrate, the metal-polymer adhesion issue. As the adhesion of metal with plastic is much weaker that of metal with silicon, the metal film on plastic substrate is easier to peal off, especially when the liquid sample is kept on the substrate for fairly long. Long time incubation is often indispensable for biological applications such as protein functionalization. Fortunately, we found the peeling off most probably happens when the liquid get to the edge of the substrate where the interface of plastic and metal is exposed, so it can be prevented by sealing the device edge.

To optimize SERS, we will try to tailor the plasmon resonance of the plastic pyramids substrates to match the excitation wavelength with the methods including using media with different refractive index, changing the size and layout of the pyramids by designing different templates or by bending and stretching of plastic substrates. To completely prevent gold peeling off from the polymer substrate, we will try to find a way to improve the adhesion of metal and plastic.

## 4. Conclusion
We demonstrated a low cost, high throughput and convenient process to make both positive and inverted polymer pyramids substrates by replica molding for the purpose of SERS. We measured and compared the SERS enhancement factor on both substrates and found the inverted pyramids substrate works much better than the positive pyramids array for SERS. Based on optical reflection measurement and FDTD simulation result, we propose two explanations for the enhancement factor difference, the plasmon resonance matching and SERS "hot spots" distribution. In the end we discussed the advantages and limitations of this technique and proposed the future directions.

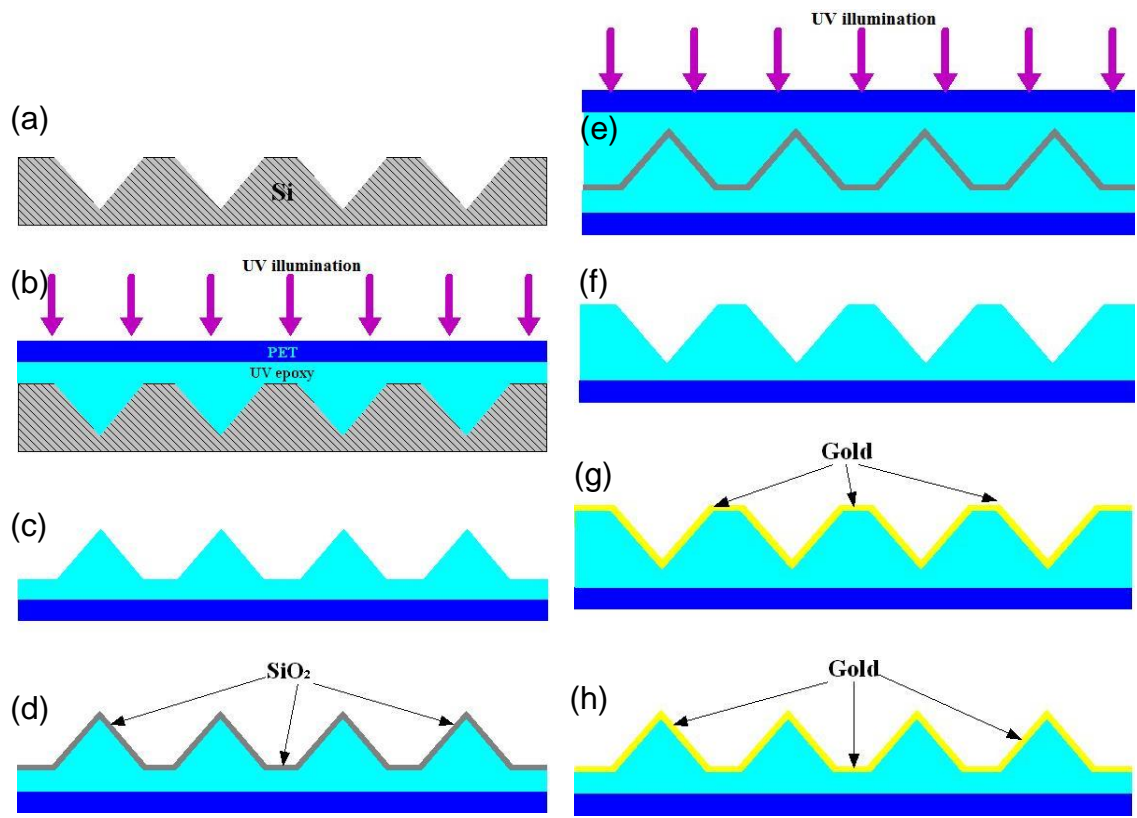

Figure.1. Fabrication process of positive and inverted pyramids replica SERS substrates. (a)Inverted pyramids silicon template. (b) Polymer molding on silicon master and cured by UV illumination.(c)Positive pyramids replica after peeled off.(d)Positive pyramids template made by e-beam evaporation of 20nm SiO2 onto positive pyramids replica.(e)Polymer molding on the positive pyramids template and cured by UV illumination.(f)Inverted pyramids replica after peeled off.(g)Inverted pyramids SERS substrate completed by deposition of 10nm of Titanium followed by 200nm of gold onto inverted pyramids replica.(f)Positive pyramids SERS substrate completed by deposition of 10nm of Titanium followed by 200nm of gold onto positive pyramids replica.

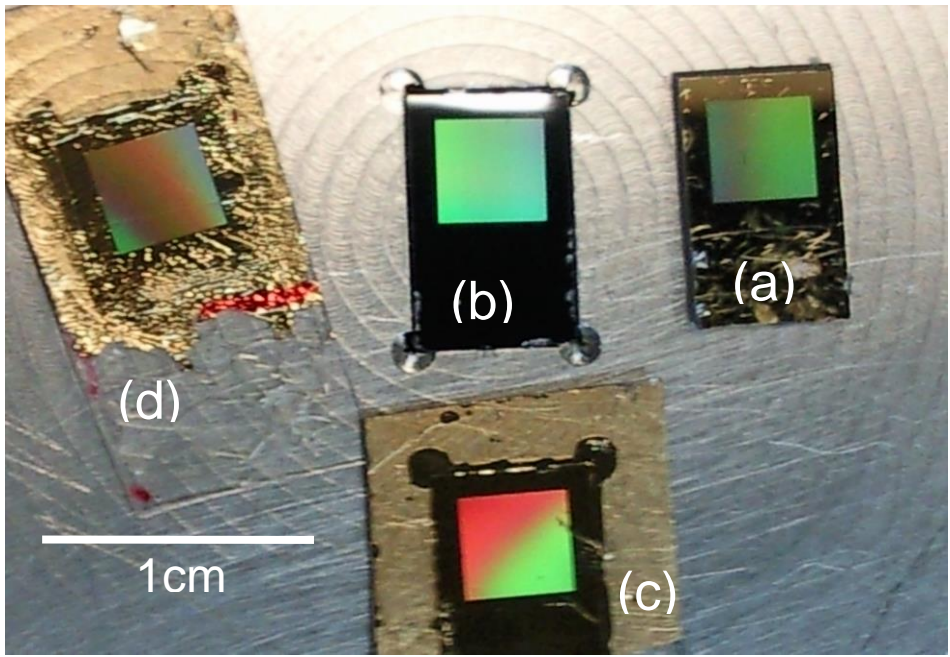

Figure.2. Photographs of (a)Klarite SERS substrate. (b)Klarite SERS substrate with gold coating removed as inverted pyramids silicon template. (c)Completed positive pyramids replica SERS substrate with gold coating. (d)Completed inverted pyramids replica SERS substrate with gold coating.

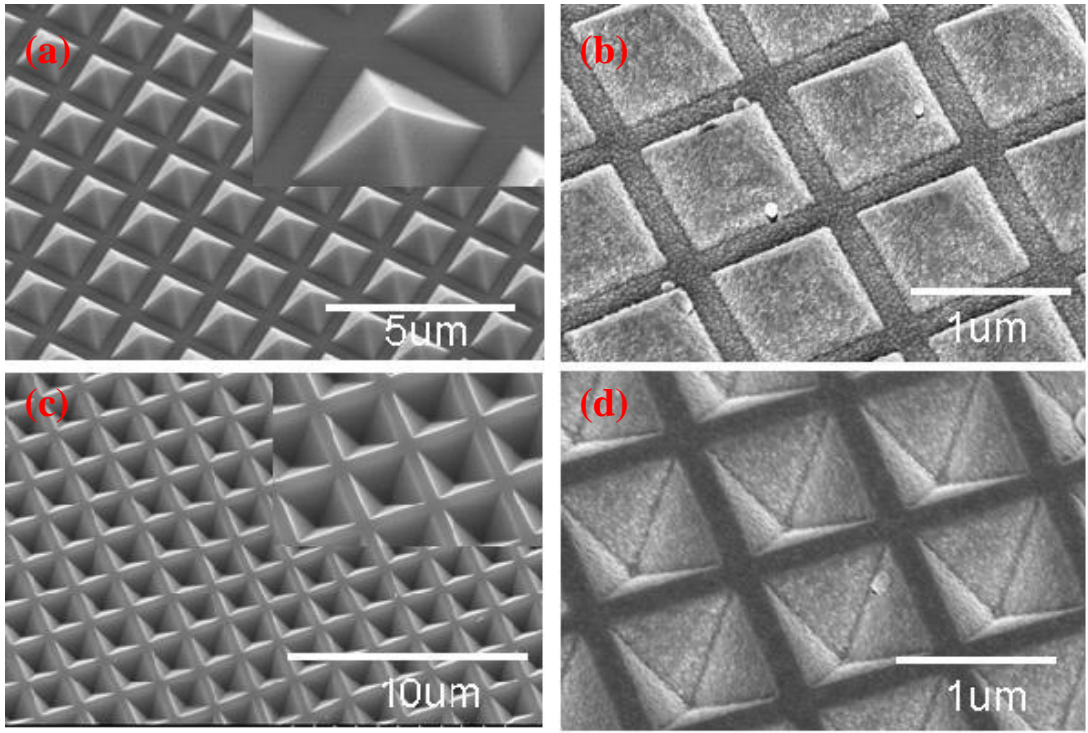

Figure 3. SEM of pyramids replica.(a)Inverse pyramids replica. The inset is zoomed-in image.(b)Inverse pyramids replica with 200nm gold deposited. (a)Positive pyramids replica. The inset is zoomed-in image.(b)Positive pyramids replica with 200nm gold deposited.

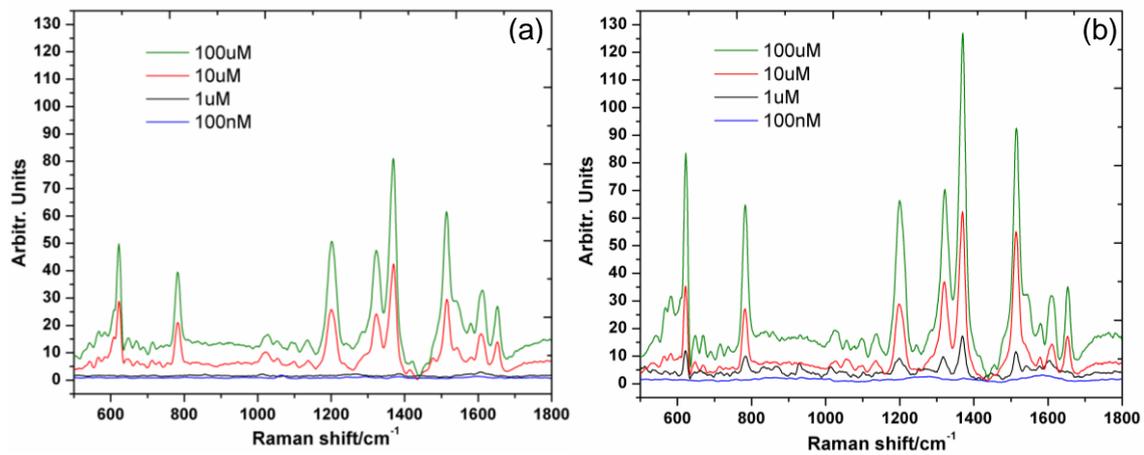

Figure 4. Raman spectra of R6g with different concentration(100uM, 10uM, 1uM, 100nM) on positive pyramids replica SERS substrate(a) and inverted pyramids replica substrate(b).

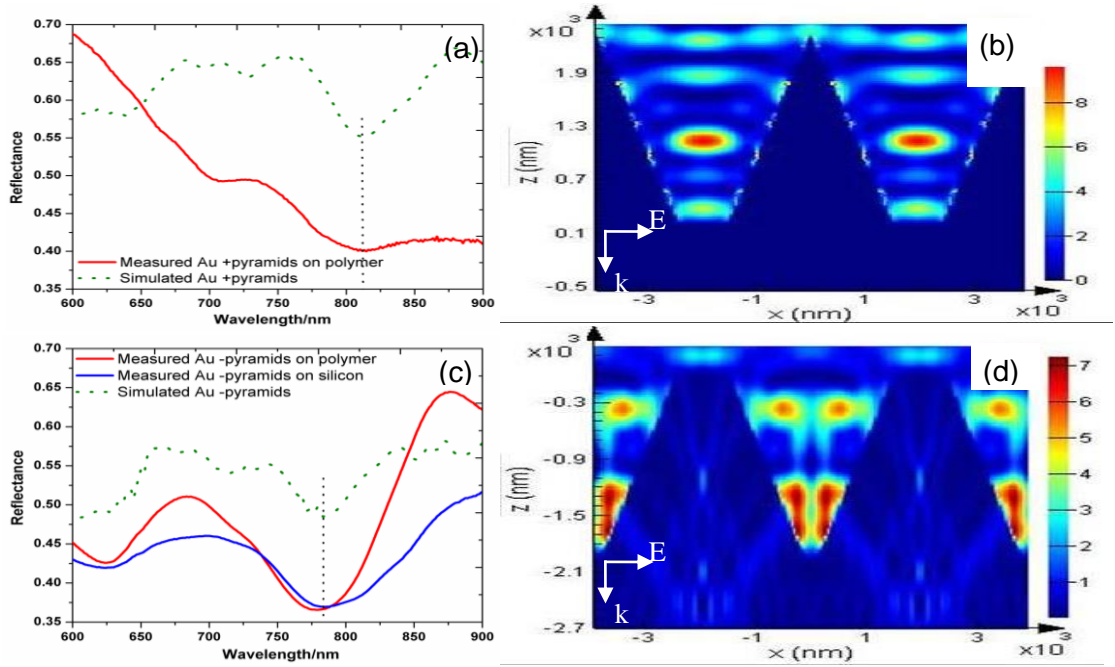

Figure 5. Spectroscopic reflection measurement and 3D FDTD simulation of positive and inverted pyramids.with 200nm gold coating. (a)Measured reflection spectra(red solid curve) and simulated reflection spectra(green dotted curve) of positive pyramids replica SERS substrate.(b)Normalized scattered electric field distribution on positive pyramids at the excitation wavelength of 785nm. (c) Measured reflection spectra(solid curves) and simulated reflection spectra(green dotted curve) of inverted pyramids SERS substrate. (d) Normalized scattered electric field distribution on inverted pyramids at the excitation wavelength of 785nm.